\documentclass[showpacs,preprintnumbers,amsmath,amssymb]{revtex4}
\usepackage{graphicx}
\usepackage{dcolumn}
\makeatletter
\input{epsf}
\begin{document}
\title{Relation between concurrence and Berry phase of an entangled state of two spin 1/2 particles }
\author{Banasri Basu}
 \email{banasri@isical.ac.in}
\affiliation{Physics and Applied Mathematics Unit\\
 Indian Statistical Institute\\
 Kolkata-700108 }

\begin{abstract}
We have studied here the influence of the Berry phase generated due to a cyclic evolution of an
entangled state of two spin $1/2$ particles. It is shown that the measure of formation of entanglement is related to the cyclic geometric phase of the individual spins.\\ \\

\end{abstract}
 
\maketitle

In the last few years Berry phase in a single particle
system has been studied very well, both theoretically and experimentally.
However, the study of Berry phase in an entangled state has become a topic of interest very recently \cite{1,2,3,4,5}.  As entanglement is a striking feature of quantum systems and is a useful resource
to realize quantum information, quantum teleportation, quantum cryptography
or quantum computation, the study of entangled state has absorbed much attention.
We have to be careful about two aspects of Berry phase of an entangled state. One is how it changes due to interparticle interaction \cite{1,2,3} and the other is how it affects \cite{4} the biparticle states. The motivation of this letter is to see the effect of the Berry phase on an entangled system of two spin-1/2 particles. As a result we have also found a connection between
it and the concurrence (which is the measure of of entanglement) for different spin chains.

It is well known that a fermion which is a spin 1/2 particle can be
realized when a scalar particle is attached with a magnetic flux
quantum.  The attachment
of the magnetic flux quantum changes the spin and statistics of the
scalar particle. The effect of the flux quantum is to induce an
appropriate Aharanov- Bohm phase which simulates the statistical
phase factor. For one complete rotation the wave function of a
fermion acquires an extra geometric phase known as Berry phase
besides the dynamical phase. The attachment of magnetic flux quantum to a scalar
particle is also equivalent to
the motion of a charged particle
in the field of a magnetic monopole of strength $2b$ where $b$ is a half integer or integer.
The angular momentum of the particle can be written as
\begin{equation}
{\bf J}={\bf r}\times {\bf p}-b{\hat{\bf r}}
\end{equation}
This produces a magnetic field \cite{wo}
\begin{equation}
B=\frac{b}{2\pi R^2}\phi_0
\end{equation}
 normal to the surface
 where $\phi_0=\frac{hc}{e}$ is the flux quantum and R is the radius of the sphere.
For one complete
rotation, wave function of the particle will acquire an extra geometric phase as \cite{dbpb}
\begin{equation}\label{phase}
e^{i\gamma}=~ e^{i 2 \pi
~b}
\end{equation}
besides the dynamical phase. $b$ is called the Berry phase factor and in this letter we shall show how this factor is related to the concurrence of an entangled state. As $b=1/2$ corresponds to one flux
quantum, when a scalar field traverses a closed path with one flux
quantum, we have the phase as $e^{i2\pi b}=e^{i\pi}$. In general, the Berry phase acquired by a fermion when it encircles $N$ number of magnetic flux quanta in
a closed path is given by $e^{i2\pi  N}$.

To study the Berry phase effect of an entangled system of two
spin-1/2 particles (at A and B) we may start with the state describing the two spins in the standard basis as
\begin{equation}\label{Bel}
\psi=\frac{1}{\sqrt M}[a_1|\downarrow \downarrow>+ a_2 |\downarrow\uparrow>+a_3|\uparrow\downarrow>
+a_4|\uparrow\uparrow>]
\end{equation}
where $a_{i}$'s are the complex coefficients and $M=\sum_{i=1}^4|a_i|^2$. We have to be
careful about the influence of one particle on the other. The
magnetic flux line of a said fermion will change its direction due
to the presence of the other. This may
be considered as the magnetic field ${\bf B}$($t$) rotating with an angular
velocity $\omega$ around the $z$-axis under an arbitrary angle
$\phi$. We may consider the time dependent magnetic field
given by
\begin{equation}\label{Bt}
    {\bf B}(t)=B {\bf n}(\phi,t)
\end{equation}
when the unit vector ${\bf n}(\phi,t)$ may be depicted as

\begin{equation}\label{nn}
    {\bf n}(\phi,t)=
    \left(
    \begin{array}{cc}
    \sin \phi & \cos(\omega_0 t) \\
    \sin \phi & \sin(\omega_0 t) \\
    \cos \phi
    \end{array}\right)
\end{equation}
The field {\bf B}(t) acts as an external parameter and let us assume that the system at A is driven by the external fields.
Let us now write
the instantaneous eigenstates of a spin operator in direction
 ${\bf n}(\phi,t)$ expanded in the $\sigma_z$-basis (where $\vec{\sigma}$ are Pauli matrices) are given by
\begin{equation}\label{art}
\begin{array}{ccc}
   |\uparrow>= |\uparrow_n;t>=\cos \frac{\phi}{2} |\uparrow_z> +~ \sin
    \frac{\phi}{2} e^{i\omega t}|\downarrow_z> \\
    \\
   |\downarrow>=|\downarrow_n;t>=\sin \frac{\phi}{2} |\uparrow_z> +~ \cos
    \frac{\phi}{2} e^{i\omega t}|\downarrow_z>
\end{array}
\end{equation}

For the time  evolution from $t=0$ to $t=\tau$ where
$\tau=\displaystyle{\frac{2\pi}{\omega}}$ each eigenstate will pick
up a geometric phase (Berry phase) apart from the dynamical phase
which is of the form \cite{4}
\begin{eqnarray}\label{upn}
\displaystyle{|\uparrow_n;t=0>\rightarrow|\uparrow_n;t=\tau>~\equiv~\psi_{\uparrow}(\tau)=
e^{i\gamma_+
(\phi)}
~e^{i\theta_+} |\uparrow_n;t=0>=e^{i\gamma_+
(\phi)}
~e^{i\theta_+} \psi_{\uparrow}(0)} \nonumber \\
\displaystyle{|\downarrow_n;t=0>\rightarrow|\downarrow_n;t=\tau>~\equiv~\psi_{\downarrow}(\tau)
=e^{i\gamma_- (\phi)} ~e^{i\theta_-} |\downarrow_n;t=0>=e^{i\gamma_-
(\phi)} ~e^{i\theta_-}\psi_{\downarrow}(0)}
\end{eqnarray}
where $\gamma_\pm$ is the Berry phase which is half of the solid
angle swept out by the magnetic flux line and
$\theta_\pm$ is the dynamical phase. The dynamical phase $\theta_\pm$ can be eliminated by using the so-called $spin$-$echo$ method \cite{4} and henceforth we shall concentrate on the effect of the Berry phase. We can then write
\begin{equation}\label{ab}
\psi_{\uparrow}(\tau)=
e^{i\gamma_+
(\phi)}
 \psi_{\uparrow}(0)=e^{i\gamma_+}\psi_{\uparrow},~~~~~~\psi_{\downarrow}(\tau)
=e^{i\gamma_-
(\phi)} ~\psi_{\downarrow}(0)=e^{i\gamma_-}\psi_{\downarrow}
\end{equation}
The explicit values of the Berry phase is given by $\gamma_\pm$ as
\begin{equation}\label{gpn}
    \begin{array}{cl}
      \displaystyle{\gamma_+(\phi)}~= & \displaystyle{-\pi(1-\cos \phi)} \\
      &\\ \\
      \displaystyle{\gamma_-(\phi)}~= & \displaystyle{-\pi(1+\cos \phi) }\\
    \end{array}
\end{equation}
This result can be explained as the system
B that evolves freely has no effect on any behaviour of the system A as long as the whole system is initially prepared in a separable state.

We can now construct an entangled state in terms of eqns.(\ref{ab}). To
illustrate the effect of the Berry phase factor $\gamma_{+(-)}$ on an entangled state let us consider the state
\begin{equation}\label{wave}
\displaystyle{\psi}~=\displaystyle{
\frac{1}{\sqrt{|\alpha|^2+|\beta|^2}}(\alpha |\downarrow\downarrow>+\beta |\uparrow\uparrow>})
\end{equation}
which can be written as 
\begin{equation}
\psi_\tau= 
\displaystyle{\frac{1}{\sqrt{|\alpha|^2+|\beta|^2}}(\alpha e^{2i\gamma_-} |\psi_\downarrow\psi_\downarrow>+\beta e^{2i\gamma_+}|\psi_\uparrow\psi_\uparrow>})\\ \\
\end{equation}
after evolution of a time period  from t=0 to $\tau$.
Here we have assumed that the geometric phase is generated due to the cyclic motion of the particle A, under the influence of the particle B.
The geometric phase of this initial entangled state is controlled by the geometric phase of the subsystem only.
If we have a cyclic evolution of this entangled state the total phase generated will be different.
The Berry phase $\Gamma$ of this entangled state will be  given by \cite{5,pra}
\begin{equation}
e^{i\Gamma}=\alpha~e^{2i\gamma_-}+\beta~e^{2i\gamma_+}=
\alpha~e^{-2i\gamma_+}+\beta~e^{2i\gamma_+}
\end{equation}
The Berry phase of this entangled state can also be calculated from its definition as
\begin{equation}
\gamma_{ent}=\oint <\psi|\nabla\psi>d\phi
\end{equation}
where $\psi=\psi(t)$ is given by the eqn.(\ref{wave}) and the integration is over the angle $\phi$ which gives the deviation of the magnetic flux line from the quatization axis.

We can also construct the symmetric( antisymmetric) Bell state from the cyclic evolution (from t=0 to t=$\tau$) of one single spin state with respect to the other as
\begin{equation}
|\psi_{\pm}>_\tau~\equiv~\frac{1}{\sqrt 2}(e^{i\gamma_+}|\psi_\uparrow \psi_\downarrow >
\pm e^{-i\gamma_+}|\psi_\downarrow \psi_\uparrow >)
\end{equation}
We note that $\gamma_{\pm}$ is diagonal in $|\psi_\uparrow>$ or $|\psi_\downarrow>$ basis
but it is not diagonal in $|\psi_{\pm}>$.
If we choose
\begin{equation}
\Sigma ~=~ \left( \begin{array}{cc}
\cos \gamma_+ & - i \sin \gamma_+\\
i \sin \gamma_+ & \cos \gamma_+\\ \end{array} \right)
\end{equation}
we have
\begin{equation}
\left( \begin{array}{l}
| \psi_+>\\
| \psi_->\\ \end{array} \right)_\tau ~\equiv~ \Sigma
\left( \begin{array}{l}
| \psi_+>\\
| \psi_->\\ \end{array} \right)
\end{equation}
for the time evolution from t=0 to t=$\tau$. We can then say that $\Sigma$ is the matrix phase factor(geometric) which explains the cyclic evolution of the symmteric and antisymmetric entangled Bell state. This explains why the symmetric Bell state is not returned to the symmetric or antisymmetric Bell state after one compelte rotation.\\
We would now like to make a comment on the entanglement of formation or
the concurrence of these entangled states and show its relation with the Berry phase. For an entangled  Bell
state given by (11), the complex concurrence $C$ is given by
\cite{woot,MilM, andrea}
\begin{equation}\label{Cs}
    C=2\alpha\beta
\end{equation}
and its norm equals the standard concurrence. $C$ extracts the information about the entanglement between the two spins from the probabilities and phases relative to specific two-spin states.
Comparison of eqns. (8), (10) with eqn.(\ref{phase})
we find
\begin{equation}\label{ps}
  {\gamma_+(\phi)~\rightarrow~b}=-\frac{1}{2}(1-\cos \phi),
\end{equation}
which shows that in an entangled state the
the angle $\phi$ is dependent on the factor $b$ which produces the field {\bf B} and $\gamma_+$ depends on it. Thus the weights of the entangled state (11) i.e.  
 the coefficients $\alpha$ and $\beta$ will be the functions of the angle $\phi$ i.e.
 the deviation of the magnetic flux line from the quantization axis. It is easily understood that the change of the angle $\phi$ will change the amount of entanglement of formation or concurrence of the two-spin state.
Now let us choose the coefficients $\alpha$ and $\beta$  such that the
positive definite norms  $|\alpha|$ and $|\beta|$ are given by
\begin{equation}
\frac{1}{\sqrt{2}}\left(
\begin{array}{c}
|\alpha|\\
|\beta|
\end{array}
\right) =\left(
\begin{array}{c}
\cos^2\frac{\phi}{4}\\
\sin^2\frac{\phi}{4}
\end{array}
\right)
\end{equation}
Then the concurrence is  \begin{equation}
C(\phi)=2|\alpha|~|\beta|=\sin^2\frac{\phi}{2}=\frac{1}{2}(1-\cos\phi)=|b|
\end{equation}
 Thus it is found that the norm of the complex concurrence of
an entangled state is related with its
 Berry phase. The relation is given by
 \begin{equation}\label{conc}
C=|b|=\frac{1}{2}(1-\cos \phi) \end{equation}
The Berry phase factor $\mid b\mid =\frac{1}{2}(1
-\cos \phi)$ gives the measure of formation of entanglement which
is usually given by the concurrence.
The comparison of this result with eqn.(\ref{ps}) shows that the cyclic geometric phase of the entangled state becomes nontrivial \cite{pra}, it is dependent on the phase factor of
the single spin only.
But this result helps us to study pairwise entanglement in Heisenberg spin models.
It follows that the measure of
entanglement between two states in a multi-spin state is dependent on the angle $\phi$
which may vary with the change of temperature and/or the external magnetic
field on the spin system.
We may  mention here that entanglement in a state has been computed
both at zero temperature and finite temperature (thermal
entanglement) for a variety of spin models \cite{ind}.
Before proceeding further a comment regarding the value of $|b|$ is pertinent. As $|b|$
depends on the angle $\phi$,
$|b|$ lies within the range $0\leq |b| \leq 1$,
but all the continuous values of $\phi$ will not correspond to the physical fixed
values of $|b|$. We have to be careful about this point.
It is easy to see that for some known values of the angle $\phi$ we arrive at
different spin models and can also achieve the correct measure of entanglement $C$.\\
 (i) For $\phi=0$ (i.e.
magnetic flux line is along the $z$-axis)\\
we have
\begin{equation}
C=|b|=0
\end{equation}
 indicating that the system is
disentangled. This is the case for a isotropic ferromagnet where it is found
that there is no entanglement at any temperature or magnetic field
strength \cite{prat}.

(ii) For the angle $\phi =\pi/2$ we arrive at the next physical fixed
value of $|b|=1/2$ . \\ It is
noted that in this case the spin axis and the direction of the
magnetic flux is orthogonal to each other. This happens in a
frustrated spin system of an antiferromagnetic chain. Indeed a
frustrated spin system is characterized by a resonating valence bond
(RVB)pair of singlets \cite{pwan} and we note that all nearest neighbour
pairs are entangled with \begin{equation}
C=|b|=1/2
\end{equation}
The frustration in
an antiferromagnetic system may occur either due to the topology of
the spin lattice such as a triangular lattice or by the next nearest
neighbour interaction (NNN) and the resonating valence bond is
formed by singlets of nearest neighbour spin pairs.

(iii) For $\phi=\pi$, we get the next physical fixed value of $|b|$.
In this spin singlet system
the concurrence is given by
\begin{equation}
C=|b|=1
\end{equation}
 which
denotes the maximum entanglement .

If we can calculate the Berry phase of an entangled system we can infer about the concurrence of that particular spin system. The plot of C versus $\phi$ may show the variation of the concurrence with the angle between the quantization axis and the direction of the magnetic flux line attached with the spinor. This feature also helps us to reveal the critical behaviour of spin models \cite{oes,OsN,ent}

In the same method we can also study the Berry phase effect on the three particle entangled state of spin-1/2 particles.
Let us assume that there are three spin-1/2 particles in a chain at three equidistant points A, B and C.
The Berry phase of the 3-particle entangled state will be given by
\begin{equation}
e^{i\gamma(ABC)}=a_1e^{i\gamma(AB)}e^{i\gamma(C)}+a_2e^{i\gamma(A)}e^{i\gamma(BC)}
+a_3e^{i\gamma(B)}e^{i\gamma(CA)}
\end{equation}
With the help of the Berry phase of the two particle states, already known, and from
  the knowledge of relative probabilities  $a_i$'s we can calculate $\gamma(ABC)$. This will give us the measure of entanglement or the concurrence of this 3-spin system.
 If the spins at A and B are maximally entangled then the entanglement of AC is zero.
The maximality of entanglement
is a specific factor of quantum phase transition and at this
critical point correlations develop on all length scales and in some
sense at this point the state is delocalized and the spin pair local
entanglement is shared by all the spins.

 In a many body system the existence of
a quantum phase transition strongly influences the behavior of the
system near the critical point with the development of long range
correlation. It is expected that entanglement is responsible for the
existence of such correlation \cite{OsN}. The first order quantum
phase transitions can bring about macroscopic changes in the amount
of pairwise entanglement in spin systems. To study the role of entanglement
in a critical phenomenon \cite{oes}, in a
spin system, we consider the nearest neighbour (NN) spin pairs and
the associated concurrence with each pair.
We may now consider the entanglement between one spin and the rest
$(n-1)$ spins in the total lattice. It is expected that the total
one-versus-rest entanglement is larger than the individual sum of
the two-spin nearest neighbour entanglements \cite{vedr}. This
suggests the inequality \cite{kundu}
\begin{equation}\label{nt}
    (n-1)C_{12} \leq C_{1(2,3,.....,n)}
\end{equation}

or,
\begin{equation}\label{int}
   C_{12}\leq \frac{C_{1(2,3,....,n)}}{(n-1)}\leq\frac{1}{n-1}
\end{equation}
where the second inequality follows from the fact that the maximum
entanglement between the one spin and the rest is equal to $1$.
Indeed, apart from one spin, the rest of the spin system may be
viewed as a block spin in terms of the block variable
renormalization scheme. At the critical point we will have the
maximum entanglement so that at this point
\begin{equation}\label{e12}
C_{12} = \frac{C_{1(2,3,....,n)}}{(n-1)}=\frac{1}{n-1}
\end{equation}

This implies that the two-spin entanglement is now shared by all
spins and the bipartite entanglement decreases as
$\frac{1}{n-1}\approx \frac{1}{n}$ for large $n$. In the
thermodynamic limit the bipartite entanglement vanishes  and the
entanglement that survives is the entanglement of the total lattice.
 The sharing of the bipartite entanglement is responsible for the
 long range correlation which develops at the
critical point.\\ \\

Acknowledgement: The author thanks the referee for his good suggestions and constructive comments. The author is also grateful to Prof. P. Bandyopadhyay and Dr. D. Banerjee for helpful discussion.

\end{document}